\documentclass[12pt]{article}

\usepackage{graphicx} 

\begin{document}





\title{The relation between frequentist confidence intervals  and Bayesian credible intervals}

\author{S.I.Bitioukov and N.V.Krasnikov
\\
INR RAS, Moscow  117312}

\maketitle

\begin{abstract}
We investigate the relation between frequentist and Bayesian approaches. Namely, we 
find the ``frequentist'' Bayes prior
$\pi_{f}(\lambda,x_{obs}) = -\frac{  \int_{-\infty}^{x_{obs}}\frac{\partial f(x,\lambda)}{\partial \lambda  }dx}{f(x_{obs},\lambda)}$ 
(here $f(x,\lambda)$ is the probability density)
for which the results of frequentist and Bayes approaches to 
the determination of confidence intervals coincide. In many cases 
(but not always)
the ``frequentist'' prior which reproduces frequentist results coincides with the Jeffreys prior.

\end{abstract}

\newpage

One of the standard problems in statistics \cite{1} is an estimation of  the values of unknown 
parameters in the probability density.  
There are two methods to solve this problem - the frequentist and the Bayesian.

In this paper we investigate the relation between frequentist and Bayesian approaches. Namely, we 
find the ``frequentist'' Bayes prior
$\pi_{f}(\lambda,x_{obs}) = -\frac{  \int_{-\infty}^{x_{obs}}\frac{\partial f(x, \lambda)}{\partial \lambda  }dx}
{f(x_{obs},\lambda)}$ 
(here $f(x,\lambda)$ is the probability density)
for which the results of frequentist and Bayes approaches to 
the determination of confidence intervals coincide. In many cases 
(but not always)
the ``frequentist'' prior  coincides with the Jeffreys prior. Note that in ref.\cite{2}
the relation between frequentist confidence intervals and Bayesian credible intervals has been 
found for probabilities densities of the special type $f(x,\lambda) = \Phi(x - \lambda)$ and 
$f(x,\lambda) = \frac{1}{\lambda}F(\frac{x}{\lambda})$.

As an example consider the case of 
random  continuous observable $-\infty < x  < \infty $ with the probability density 
$f(x, \lambda)$\footnote{Here $\infty <\lambda < -\infty $ is some 
unknown parameter and $\int^{\infty}_{-\infty}
f(x,\lambda)dx = 1$.}. 

In Bayesian method  \cite{1,3} due to Bayes theorem 
\begin{equation}
P(A|B) = \frac{P(B|A)P(A)}{P(B)}
\end{equation}
the probability density for unknown parameter  $ \lambda $  
is determined as
\begin{equation}
p(\lambda |x_{obs}) = \frac{f(x_{obs},\lambda)\pi(\lambda)}{\int_{-\infty}^{+\infty} 
f(x_{obs},\lambda^{'})
\pi(\lambda^{'})d \lambda^{'}} \,.
\end{equation}
Here $x_{obs}$ is the observed value of the random variable  $x$ and 
 $\pi(\lambda)$ is the prior function. In general 
the prior function   $\pi(\lambda)$    is not known that is the 
main problem of the Bayesian approach. Formula (2) reduces the statistics problem to the 
probability problem.  
The probability that parameter $\lambda$
lies in the interval $ \lambda_{down} \leq \lambda \leq \lambda_{up}$ is 
\footnote{Usually 
$\alpha(\lambda_{up},\lambda_{down}|x_{obs})$ is taken nondependent on $\lambda_{up}$, $\lambda_{down}$ and equal to $0.05$.}
\begin{equation}
P(\lambda_{down} \leq \lambda \leq \lambda_{up}) =
\int_{\lambda_{down}}^{\lambda_{up}} p(\lambda|x_{obs}) d \lambda 
= 1 - \alpha(\lambda_{up},\lambda_{down}|x_{obs}) \,,
\end{equation}
where
\begin{equation}
\alpha(\lambda_{up},\lambda_{down}|x_{obs}) = \beta(\lambda_{up}|x_{obs}) + \gamma(\lambda_{down}|x_{obs})\,,
\end{equation}
\begin{equation}
\beta(\lambda_{up}|x_{obs}) = \int^{\infty}_{\lambda_{up}}
p(\lambda|x_{obs})dx \,,
\end{equation}
\begin{equation}
\gamma(\lambda_{down}|x_{obs}) = \int_{-\infty}^{\lambda_{down}}p(\lambda|x_{obs})dx = 
1 - \beta(\lambda_{down}| x_{obs})\,,
\end{equation}

The solution of the equation (3) is not unique. The most popular are the following options \cite{1}:

1. $\lambda_{down} = -\infty $ - upper limit.

2. $\lambda_{up} = \infty$ - lower limit.

3. $\int_{-\infty}^{\lambda_{down}} p(\lambda|x_{obs}) d \lambda = 
\int_{\lambda_{up}}^{\infty} p(\lambda|x_{obs}) d \lambda = \frac{\alpha}{2}$ - symmetric interval.

4. The shortest interval -  $p(\lambda|x_{obs})$ inside the interval is bigger or equal to  
$p(\lambda|x_{obs})$ outside the interval.

In frequentist approach the Neyman belt construction \cite{4}(see Fig.~1)
 is used for the determination 
of the confidence intervals.

\begin{figure}[!Hhtb]
\includegraphics[width=0.90\textwidth]{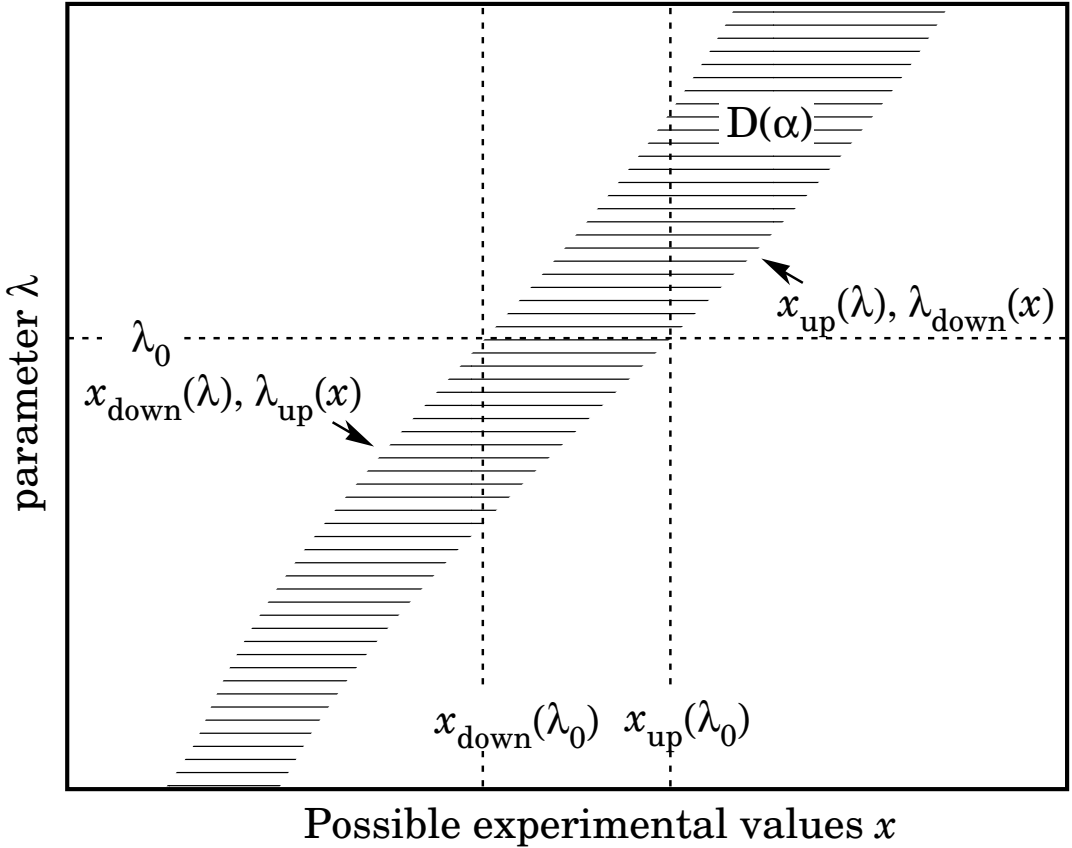}   
\caption{Neyman belt construction}
    \label{fig:1} 
\end{figure}


Namely, we require that
\begin{equation}
\int^{x_{up}(\lambda)}_{x_{down}(\lambda)}f(x,\lambda)dx = 1 - \alpha\,,
\end{equation}
or
\begin{equation}
\int^{ \infty}_{x_{up}(\lambda)}f(x,\lambda)dx + \int_{-\infty}^{x_{down}(\lambda)}f(x,\lambda)dx = \alpha \,.
\end{equation}
It should be stressed  that in general case both  $\int^{ \infty}_{x_{up}(\lambda)}f(x,\lambda)dx$ and
$\int_{-\infty}^{x_{down}(\lambda)}f(x,\lambda)dx$ can depend on $\lambda$ but their sum 
  does not depend on $\lambda$ (see eq.(8)).

The Neyman equations for the determination of upper and lower limits $\lambda_{up}$ and $\lambda_{down}$ on 
parameter $\lambda$ have the form 
\begin{equation}
\int^{x_{up}(\lambda_{up})}_{x_{obs}}f(x,\lambda_{up})dx = 1 - \alpha\,,
\end{equation}
\begin{equation}
\int^{x_{obs}}_{x_{down}(\lambda_{down})}f(x,\lambda_{down})dx = 1 - \alpha \,
\end{equation}
and they determine the confidence interval 
of possible values 
\begin{equation}
\lambda_{down} \leq \lambda \leq \lambda_{up}
\end{equation}
of the 
parameter $\lambda$ at the $(1 - \alpha)$ confidence level. 




In this paper we shall consider the   case when $\int^{ \infty}_{x_{up}(\lambda)}f(x,\lambda)dx$ does not depend on $\lambda$.
For instance, the options $x_{up}(\lambda) = \infty$,   $x_{down}(\lambda) = - \infty$ and 
$\int^{ \infty}_{x_{up}(\lambda)}f(x,\lambda)dx =  \int_{-\infty}^{x_{down}(\lambda)}f(x,\lambda)dx = 
\frac{\alpha}{2}$ correspond to the cases of upper limit on $\lambda$, lower limit on $\lambda$ and symmetric interval
correspondingly. 
As a consequence of the eqs.(8-11) and our assumption on nondependence of $\int^{ \infty}_{x_{up}(\lambda)}f(x,\lambda)dx$ 
on $\lambda$ we find that 
\begin{equation} 
\alpha = \int_{-\infty}^{x_{obs}}f(x,\lambda_{up})dx  + \int^{\infty}_{x_{obs}}f(x,\lambda_{down})dx \,.
\end{equation}  
To find the relation beetween frequentist and Bayesian approaches we have to find the prior for which 
the formulae (4-6) and (12) coincide.  Namely, we require 
that 
\begin{equation}
\int_{-\infty}^{\lambda_{down}}p(\lambda|x_{obs})d\lambda +\int^{\infty}_{\lambda_{up}}p(\lambda|x_{obs})d\lambda 
= \int_{-\infty}^{x_{obs}}f(x,\lambda_{up})dx  + \int^{\infty}_{x_{obs}}f(x,\lambda_{down})dx \,.
\end{equation}
The solution of eq.(13) is
\begin{equation}
p(\lambda| x_{obs}) = -\int_{-\infty}^{x_{obs}}\frac{\partial f(x, \lambda)}{\partial \lambda}dx  \,.
\end{equation}
Formulae (13,14) demonstrate the equivalence of the frequentist approach and the 
Bayes approach with the prior function
\begin{equation}
\pi_{f}(\lambda| x_{obs}) = 
- \frac{  \int_{-\infty}^{x_{obs}}\frac{\partial f(x, \lambda)}{\partial \lambda}dx}{f(x_{obs},\lambda)} \,.
\end{equation}

Note that in the limit $\lambda_{down} \rightarrow -\infty$ and $\lambda_{up} \rightarrow  \infty$ 
 full probability must be equal to one, namely
\begin{equation}
\lim_{\lambda_{down}\rightarrow -\infty,\lambda_{up}\rightarrow \infty}\int_{-\infty}^{x_{obs}}[f(x,\lambda_{down}) - 
f(x,\lambda_{up})]dx  = 1 \,.
\end{equation}
As a consequence of the equation (16) we find that
\begin{equation}
\lim_{\lambda_{down} \rightarrow -\infty} \int_{-\infty}^{x_{obs}}f(x,\lambda_{down})dx = 1\,,
\end{equation}
\begin{equation}
\lim_{\lambda_{up} \rightarrow \infty}\int^{x_{obs}}_{-\infty}f(x,\lambda_{up})dx = 0\,.
\end{equation}

Consider several examples. For the probability density 
\begin{equation}
f(x,\lambda) = \Phi(x - \lambda)
\end{equation}
as a 
consequence of the formulae (14-15) we find that    
\begin{equation}
p_{f}(\lambda, x_{obs}) = \Phi(x_{obs} - \lambda)\,,
\end{equation}
\begin{equation} 
\pi_{f}(\lambda, x_{obs}) = 1\,.
\end{equation}
Note that for  the distribution (19) the Jeffreys prior \cite{5} $ \pi(\lambda) \sim 
\sqrt{ [\int^{\infty}_{-\infty}  
f(x,\lambda)(\frac{\partial ln(f(x,\lambda))}{\partial \lambda})^2 dx]}  \sim const  $  
does not depend on $\lambda$, i.e. for the distribution (19) 
the frequentist approach is equivalent to the Bayes approach with
the Jeffreys prior $\pi(\lambda) = const$ \cite{2}.

Consider the case with  the parameter $\lambda \geq b$. Here $b$ is some fixed number
\footnote{In ref. \cite{6}
normal distribution $N(x,\mu,\sigma^2)$ with additional constraint $\mu \geq 0$ has been studied.}. 
Such situation arises when we measure signal $s$ in the presence of nonzero background $b$ and $\lambda = b + s $, $b \geq 0$, 
$s \geq 0 $.
The parameter $\lambda$ lies in the interval $b \leq \lambda < \infty $. The direct use 
of the formulae (12,13) leads to the inconsistency. Namely, we find that the  probability 
\begin{equation}
P( b < \lambda < \infty) = \int^{\infty}_{b} \Phi(x_{obs}- \lambda) d\lambda < 1
\end{equation}
that contradicts to the postulate that the full probability 
must be equal to 1. At the frequentist language the inequality (22) is the consequence of the fact that
\begin{equation}
\int^{\infty}_{x_{obs}}f(x,\lambda_{down} = b)dx \neq 0 \,.
\end{equation}
To obtain the correct solution we must use the language of the conditional probabilities. Really, the probability that 
parameter $ \lambda $ lies in the interval $ \lambda_0 \leq \lambda \leq \lambda_0 +d\lambda $ is equal to 
$ \Phi(x_{obs} -\lambda_0)d\lambda $. The probability that parameter    $ \lambda $ lies in the interval 
$ \lambda_0 \leq \lambda \leq \lambda_0 +d\lambda $ provided $\lambda \geq b$ is determined by the formula of the 
conditional probability
\begin{equation} 
P(\lambda_0 \leq \lambda \leq \lambda_0 +d\lambda| \lambda \geq b) = 
\frac{P(\lambda_0 \leq \lambda \leq \lambda_0 +d\lambda)}{P(\lambda \geq b)} = 
\frac{\Phi(x_{obs} -\lambda_0)d\lambda}{\int_{b}^{\infty}\Phi(x_{obs}-\lambda)d\lambda}  \,.
\end{equation}
So we see that  condition $\lambda \geq b$ leads to the appearance of additional 
factor $\int_{b}^{\infty}\Phi(x_{obs}-\lambda)d\lambda$ in the denominator of the formula (24). This factor 
restores  the requirement that full probability  $ P(b_0 \leq \lambda < \infty) = 1 $.  
For instance, the probability that signal  $s$  is less than $s_0$ is determined by the formula
\begin{equation} 
P(s \leq s_0)  = \frac{\int_{0}^{s_0}\Phi(x_{obs}- b -s)ds}
{\int_{b}^{\infty}\Phi(x_{obs}-\lambda)d\lambda}   
\end{equation}
and it  coincides with the corresponding formula of the $CL_s$ method \cite{7,8}.

For the probability density
\begin{equation}
F(x,\lambda) = \frac{1}{\lambda}\Phi(\frac{x}{\lambda})
\end{equation}
we find that
\begin{equation}
p_{f}(\lambda, x_{obs}) =  \frac{x_{obs}}{\lambda^{2}} \Phi(\frac{x_{obs}}{\lambda})\,,
\end{equation}
\begin{equation} 
\pi_{f}(\lambda, x_{obs}) = \frac{x_{obs}}{\lambda}\,.
\end{equation}
Again in this case the prior (28) coincides with the Jeffreys prior \cite{2}.

Consider  the probability density\footnote{For the probability density (29) $\int^{\infty}_{-\infty}f(x,\lambda)dx =
\sum_{n=0}^{n =\infty}P(n, \lambda) = 1$.}
\begin{equation}
f(x, \lambda) = \theta(x)P([x], \lambda)\,,
\end{equation}
where
\begin{equation}
P([x],\lambda) = \frac{\lambda^{[x]}}{[x]!}e^{-\lambda}
\end{equation}
is the Poisson distribution and $[x]$ is an integer part of $x$ (for instance, $[2.33] = 2$). 
Using formulae (14,15) one can find that
for 
the probability density (29)   and $x_{obs} \geq 1$\footnote{For $x_{obs} <1$ $p_f(x_{obs}, \lambda) = P(0,\lambda)$ and
$\pi_f(\lambda,x_{obs}) = 1$.} 
\begin{equation}
p_{f}(x_{obs}, \lambda) = P([x_{obs}]-1,\lambda) + (x_{obs} - [x_{obs}])(-P([x_{obs}]-1,\lambda) +
P([x_{obs}],\lambda)) \,,
\end{equation}
\begin{equation}
\pi_{f}(\lambda, x_{obs} ) = \frac{[x_{obs}]}{\lambda} + (x_{obs} - [x_{obs}])( -\frac{[x_{obs}]}{\lambda} + 1)\,.
\end{equation}
The Jeffreys prior for the distribution function (29) coincides with the Jeffreys prior for Poisson distribution 
and it is proportional to $\pi(\lambda) \sim \frac{1}{{\sqrt{\lambda}}}$. So we see that for the probability density (29)
the ``frequentist prior'' (32) and the Jeffreys prior are different. 

Consider now the case when  $\int^{ \infty}_{x_{up}(\lambda)}f(x,\lambda)dx$ depends on $\lambda$. 
Instead of $x_{up}(\lambda)$ and $x_{down}(\lambda) $ we can find another  $x_{up}^`(\lambda)$ and 
$x_{down}^`(\lambda$ such that $x_{up}(\lambda_{up}) =x_{up}^`(\lambda_{up}) = x_{obs}$, 
$x_{down}(\lambda_{down}) =x_{down}^`(\lambda_{down}) = x_{obs}$ and 
\begin{equation} 
\int_{ -\infty}^{x_{up}^{`}(\lambda)}f(x,\lambda)dx = \int_{ \infty}^{x_{obs}}f(x,\lambda_{up})dx \,,
\end{equation}
\begin{equation}
\int^{ \infty}_{x_{down}^{`}(\lambda)}f(x,\lambda)dx = \int^{ \infty}_{x_{obs}}f(x,\lambda_{down})dx \,.
\end{equation}
For the confidence intervals defined by new trajectories $x_{up}^{`}(\lambda)$ and $x_{down}^{`}(\lambda)$ the Neyman equations 
have the form
\begin{equation}
\int^{x_{up}^{`}(\lambda)}_{x^{`}_{down}(\lambda)}f(x,\lambda)dx = 1 - \alpha^{'} \,,
\end{equation}
\begin{equation}
\alpha^` =  \int_{ -\infty}^{x_{obs}}f(x,\lambda_{up})dx + 
\int^{ \infty}_{x_{obs}}f(x,\lambda_{down})dx \,.
\end{equation}
and the equations (14,15) are valid.
So we see that for the Neyman belt construction with 
$x_{up}^`(\lambda)$, $x_{down}^`(\lambda)$ and the integral  
$\int ^{\infty}_{x_{down}^{`}(\lambda)} f(x,\lambda)dx$  nondependent 
on $\lambda$ the frequentist approach with 
$\alpha^`$ given by the exppression (36) and the Bayesian approach with prior function (14) 
coincide.


In conclusion let us formulate our main result.
For the particular case when $\int^{ \infty}_{x_{up}(\lambda)}f(x,\lambda)dx$ 
does not depend on $\lambda$ we have found the ``frequentist'' Bayes prior
$\pi_{f}(\lambda, x_{obs}) = 
-\frac{  \int_{-\infty}^{x_{obs}}\frac{\partial f(x,\lambda)}{\partial \lambda  }dx}
{f(x_{obs},\lambda)}$ for which the results of the frequentist and the Bayes approaches to 
the determination of confidence intervals coincide. In many cases 
(but not always) the prior which reproduces frequentist results coincides with the Jeffreys prior.

 This work has been supported by RFBR grant N 10-02-00468.

\end{document}